\documentclass[%
aip,
amsmath,amssymb,
reprint
]{revtex4-1}
\usepackage{graphicx}% Include figure files
\usepackage{dcolumn}% Align table columns on decimal point
\usepackage{bm}% bold math
\usepackage{amsmath}
\usepackage{color}
\usepackage[normalem]{ulem}
\usepackage{balance}
\usepackage[T1]{fontenc}

\newcommand{\mycite}[1]{\scalebox{1.3}[1.3]{\raisebox{-0.80ex}{\cite{#1}}}}
\begin{document}

\preprint{AIP/123-QED}

\title{Absorption and photoluminescence properties of coupled plasmon--exciton (plexciton) systems }% Force line breaks with \\

\author{Yuqing Cheng}\affiliation{1 School of Mathematics and Physics, Beijing Advanced Innovation Center for Materials Genome Engineering, University of Science and Technology Beijing, Beijing 100083, China}%
\author{Mengtao Sun}\thanks{mengtaosun@ustb.edu.cn}
\affiliation{1 School of Mathematics and Physics, Beijing Advanced Innovation Center for Materials Genome Engineering, University of Science and Technology Beijing, Beijing 100083, China}%


\begin{abstract}
Plexciton is the formation of new hybridized energy states originated from the coupling between plasmon and exciton. To reveal the optical properties of both exciton and plexciton, we develop a classic oscillator model to describe the behavior of them. Particularly, the coupling case, i.e., plexciton, is investigated theoretically in detail. In strong coupling, the electromagnetically induced transparency is achieved for the absorption spectra; the splitting behaviors of the modes are carefully analyzed, and the splitting largely depends on the effective number of the electrons and the resonance coupling; the photoluminescence spectra show that the spectral shapes remain almost unchanged for weak coupling and change a lot for strong coupling; the emission intensity of the exciton is strongly enhanced by the plasmon and can reach to the order of $10^{10}$ for a general case.
We also show the comparisons between our model and the published experiments to validate its validity. This work may be useful for understanding the mechanism of the plexciton and for the development of new applications.
\end{abstract}

\maketitle

%\tableofcontents

\section{\label{sec:Introduction}Introduction}
%\textbf{\textit{Introduction}}.---
Plexciton, the interaction between plasmon and exciton, plays an important role in nanotechnology and nanoscience. Usually, a metal nanoparticle (MNP) or matal nanostructure supports plasmons due to the oscillation of the free electrons excited by the external electromagnetic field; while a semiconductor nanoparticle (SNP), quantum dot (QD), or two--dimensional (2D) materials support excitons which are the bound electron--hole pairs caused by the transitions between the discrete levels in the conduction and valence bands of the semiconductor.\cite{plexciton}
Because of the unique properties of the coupling between plasmon and exciton, plexciton has been investigated widely and attracts attention for numerous potential applications, such as quantum information processing,\cite{app1,app2,app3} ultrafast optical switching,\cite{app4,app5,app6,app7,app8,app9} lasing at nanoscale,\cite{app10,app11,app12,app13} optical nonreciprocity,\cite{app14,app15} surface catalytic reaction, \cite{YangR,CaoY} etc.

In general, according to the coupling strength, the coupling between plasmon and exciton can be divided into weak coupling and strong coupling. The weak coupling often refers to the case in which the spectral shapes of the plasmon mode and the exciton mode almost remain unchanged, but the intensities of the two modes vary; while the strong coupling often refers to the case in which the spectral shapes change evidently, especially the peaks of the modes shift at least to the order of the line widths of the modes. Actually, the strong coupling is not easy to achieve and is more valuable due to its unique characteristics such as energy splitting in tuning the modes of plexciton.
C. K. Dass et al. employed a hybrid metal--semiconductor nanostructure, i.e., a semiconductor quantum well coupled to a metallic plasmonic grating, to investigate quantum coherent dynamics.\cite{ach1} They revealed that the plexciton can reduce the nonradiative quantum coherence to the range of hundreds of femtoseconds.
P. Vasa et al. employed $J$--aggregate--metal hybrid nanostructures which exhibit strong plexciton coupling and observed the optical Stark effect.\cite{ach2} They used off--resoant ultrashort pump pulses to observe fully coherent plexciton optical nonlinearities, which is helpful to ultrafast all--optical switching.
A. E. Schlather et al. employed $J$--aggregate excitons and single plasmonic dimers to investigate the coupling between them.\cite{ach3} They reported a unique strong coupling regime for the first time, with giant Rabi splitting ranging from 230 to 400 meV.
X. Mu et al. employed first--principle calculation and finite element electromagnetic simulations to investigate the plasmon--enhanced charge transfer exciton of 2D $\mathrm{MoS_2/WS_2}$ heterostructures.\cite{MuX} Both strong and weak coupling Rabi--splitting are reported. The strong coupling exists in the high energy region and enhances the electromagnetic field, but it will change the wave function and electromagnetic field mode of the exciton; while the weak coupling, i.e., Purcell effect, can enhance the charge transfer exciton density with no change of the electromagnetic field mode and the exciton wave function, which is a better method to enhance the charge transfer exciton. The results can be applied in designing the plexciton devices.
L. Ye et al. employed a single gold nanorod (GNR) and 2D materials to reveal the plexciton coupling.\cite{ach4} They used the single--particle spectroscopy method and in situ nanomanipulation via atomic force
microscopy to investigate the scattering spectra of the same GNR before and after coupling. They demonstrated that the plexciton in the GNR--$\mathrm{WSe_2}$ system would induce plasmon resonance damping, and the coupling strength influences the damping rates. They also concluded that the damping effect is dominated by the contact layer between the GNR and 2D materials, which is useful for understanding of plasmon decay channels.
The above phenomena and achievements employing strong plexciton coupling are interesting and valuable. However, understanding the mechanism and principle of this hybrid system is significant for both fundamental investigations and the potential applications.

In this study, we present a classic oscillator model to reveal the optical properties of both the exciton and the plexciton. For simplicity, we use a SNP and a MNP to represent the exciton and the plasmon, respectively. Novelty effect and properties can be achieved especially using the coupling model, such as electromagnetically induced transparency (EIT) for the absorption and the splitting in the photoluminescence (PL) spectra.

\section{\label{sec:Model}Model}
We introduce a model that can evaluate the PL and absorption properties of individual and coupled nanoparticles (NPs, semiconductor or metal). Firstly, we consider the individual NP, where we divide the interaction process into two steps: one is the absorption step, the other is the emission step. Secondly, we consider the coupling case between the two NPs, where the coupling process is also divided into two steps: one is the absorption coupling step, the other is the emission coupling step.

\subsection{\label{sec:single}Individual}
\begin{figure}[tb]
\includegraphics[width=0.48\textwidth]{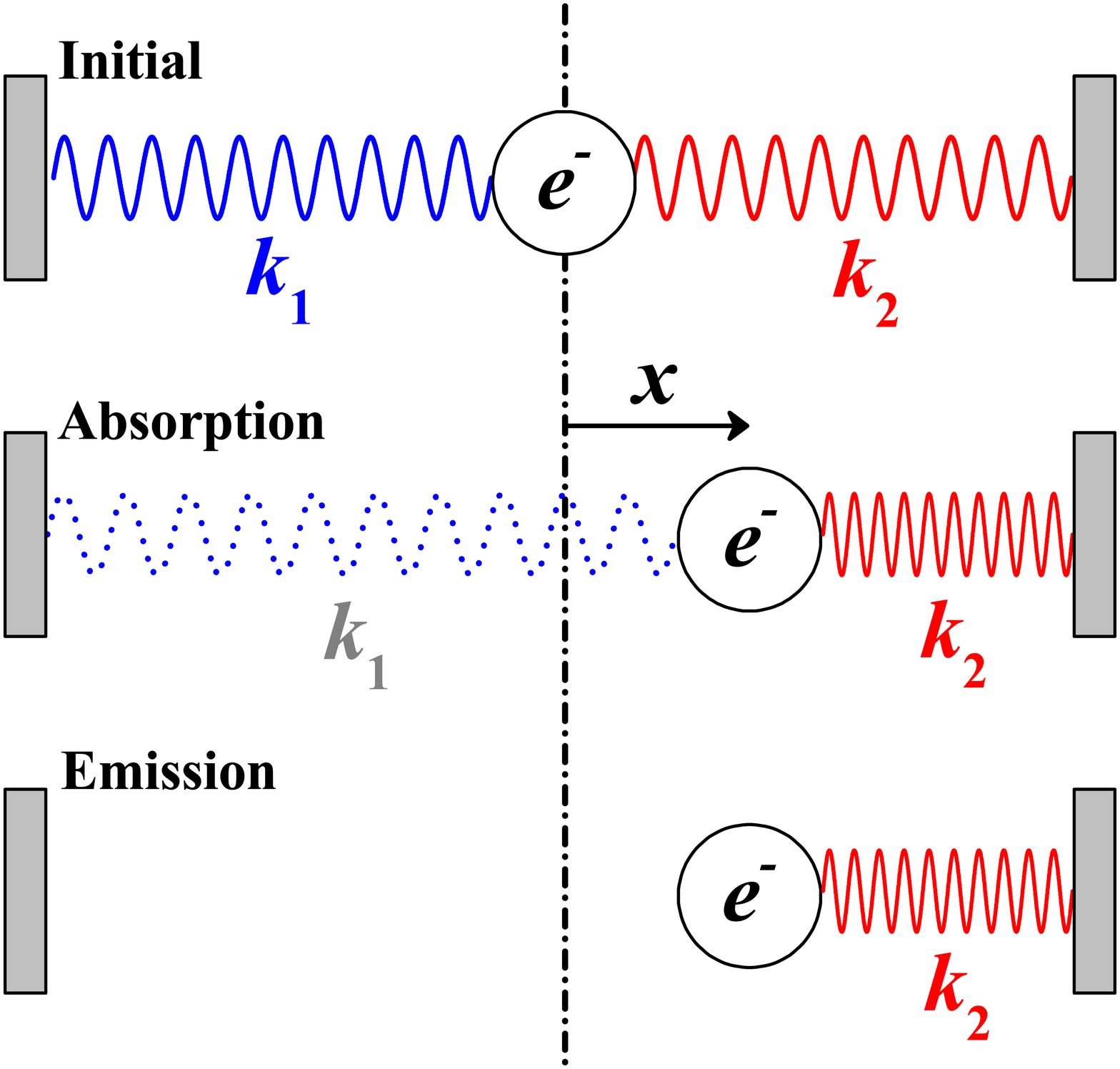}
\caption{\label{fig:scheme} Schematic of the two--step oscillator model. The blue and red curves stand for the springs with recovery factors of $k_1$ and $k_2$ respectively. The circles stand for the oscillators, which are the electrons in this model. The vertical dash--dot line stand for the equilibrium position of the oscillators. The excitation light is polarized along $x$--axis.
}
\end{figure}

%\textbf{\textit{Model}}.---
In the quantum mechanism, for semiconductor, the electrons are most in bound state, and interact with the ions much more strongly than the ones of metals do. When the electrons are excited by the incident photons, they will stay in excited state, and then the combinations of the electron--hole pairs result in the emission of photons. Refer to the above descriptions, for a SNP, we could treat the whole process in the classical view using the harmonic oscillator concept, in which we separate the process into two steps, i.e., absorption and emission, as shown in Fig. \ref{fig:scheme}. The bound electron is treated as an oscillator with certain resonance frequency due to the interaction with the ions, and there are two springs that provide the restoring. In the first step, when excited by photon, the electron absorbs the energy and start to oscillate through the interactions with the two springs; the first spring suddenly breaks as soon as the oscillator arrives to its maximum displacement, i.e., the amplitude. The broken refers to the thermal process (nonradiative process), in which part of the energy is converted into thermal energy. In the second step, the oscillator starts to oscillate through the interaction with the second spring, and emission photons simultaneously.

To start the deviation, define $x(t)$ as the displacement of an oscillator from its equilibrium position as a function of time $t$, thus $\dot{x}(t)$ and $\ddot{x}(t)$ the velocity and the acceleration. The excitation light is treated classically, with (circular) frequency $\omega_{ex}$ and electric field intensity $E_0$.

For the absorption step, the equation is in this form:
\begin{equation}
\ddot{x}+2 \beta_{a} \dot{x}+\omega_{a}^2 x=C_0 \mathrm{exp}(-\mathrm{i} \omega_{ex} t),
\label{eq:basic01}
\end{equation}
where $\omega_{a}^2=(k_1+k_2)/m_e$ is the absorption resonance frequency, $\beta_{a}$ is the absorption damping coefficient, $C_0=-e E_0/m_e$, $m_e$ is the electron mass, and $e$ is the elementary charge.
The solution is:
\begin{equation}
x(t)=A \mathrm{exp}(-\mathrm{i} \omega_{ex} t),
\label{eq:s1}
\end{equation}
with the amplitude $A=\frac{C_0}{\omega_a^2-\omega_{ex}^2-2 \mathrm{i} \beta_a \omega_{ex}}$
The energy stored in Spring 1 is:
\begin{equation}
I_a(\omega_{ex})=\frac{1}{2} k_1 |A|^2.
\label{eq:abs}
\end{equation}
When Spring 1 is broken, this energy is converted into thermal energy, i.e., the absorption spectrum could be written as $I_a(\omega)$, in which $\omega_{ex}$ is replaced by $\omega$.

For the emission step, the equation is in this form:
\begin{equation}
\begin{aligned}
\ddot{x}+2 \beta_{e} &\dot{x}+\omega_{e}^2 x=0,
\\ \mathrm{initial\ conditions:}\  &x(0)=A,\ \dot{x}(0)=0,
\end{aligned}
\label{eq:basic02}
\end{equation}
where $\omega_{e}^2=k_2/m_e$ is the emission resonance frequency, and $\beta_e$ is the emission damping coefficient. Obviously, $\omega_{a}>\omega_{e}$.
The solution is:
\begin{equation}
x(t)=\frac{A}{2\mathrm{i}\omega_s } \left[ \alpha_{+} \mathrm{exp}(\alpha_{-} t) - \alpha_{-} \mathrm{exp}(\alpha_{+} t) \right],
\label{eq:s2}
\end{equation}
where $\omega_s=\sqrt{\omega_e^2-\beta_e^2}$ is the resonance frequency of the emission, and $\alpha_{\pm}=-\beta_e \pm \mathrm{i}\omega_s$ are the conjugate eigenvalues of Eq. (\ref{eq:basic02}). The far field electric field produced by the electrons is evaluated by \cite{couple0}:
\begin{equation}
\begin{aligned}
& E(t)  \cong  \frac{N_s e}{4\pi \varepsilon_0 c^2 d} \ddot{x}(t)
\\  = &\frac{N_s e}{4\pi \varepsilon_0 c^2 d}\ \frac{A \omega_e^2}{2\mathrm{i}\omega_s }  \left[ \alpha_{-} \mathrm{exp}(\alpha_{-} t) - \alpha_{+} \mathrm{exp}(\alpha_{+} t) \right],
\end{aligned}
\label{eq:fart}
\end{equation}
where $N_s$ is the effective electron number of the SNP, $\varepsilon_0$ is the permittivity in vacuum, $c$ is the velocity of light, and $d$ is the distance between the field point and the SNP. The far field electric field in frequency domain can be evaluated by:
\begin{equation}
I_e(\omega)=\mathrm{Re}\left[\left< \int_{0}^{\infty} E^{*}(t)E(t+\tau)\mathrm{exp}(\mathrm{i}\omega \tau) \mathrm{d} \tau \right>\right],
\label{eq:ft}
\end{equation}
where $\mathrm{Re}[Q]$ is the real part of $Q$, and $<Q>$ is the time average of $Q$. Therefore, the emission or the PL spectrum is:
\begin{equation}
\begin{aligned}
I_e(\omega)&=D \left[ \frac{1}{(\omega-\omega_s)^2+\beta_e^2} + \frac{1}{(\omega+\omega_s)^2+\beta_e^2} \right]
\\ & \cong  \frac{D}{(\omega-\omega_s)^2+\beta_e^2},
\end{aligned}
\label{eq:farw}
\end{equation}
where $D=\left| A \right|^2(\frac{N_s e}{4\pi \varepsilon_0 c^2 d}\ \frac{\omega_e^2}{2 \omega_s})^2 \cdot \frac{\omega_e^2 }{2 t_0}[1-\mathrm{exp}(-2\beta_e t_0)]$. Here, we omit the second term of Eq. (\ref{eq:farw}), because the intensity of the first term is much larger than the intensity of the second term when $\omega$ is around $\omega_s$, which is corresponding to the general case of a practical PL spectrum.
Define the PL excitation (PLE) as the integrated PL intensities varying with the incident frequency, thereby, it can be evaluated by:
\begin{equation}
I_{PLE}(\omega_{ex})=\left| A \right|^2 \int_{\omega_{cut1}}^{\omega_{cut2}} \frac{\mathrm{d}\omega}{(\omega-\omega_s)^2+\beta_e^2},
\label{eq:PLE}
\end{equation}
where we retain $A$ from $D$ because the rest quantities in $D$ are constant for a certain SNP, and only $A$ depends on $\omega_{ex}$. For the practical purpose, we employ $\omega_{cut1}=\omega_{s}-2\beta_e$ and $\omega_{cut2}=\omega_{ex}$ as the lower and upper limits of the integration term.

Actually, the PL progress of a MNP can also be treated in the same way as the one of a SNP. The difference is that for MNP, $k_1 \ll k_2$ indicating that $\omega_a\approx \omega_e$ with $\omega_a$ a little larger than $\omega_e$. This treatment is equivalent to the one in our previous work where we considered only one spring rather than two.\cite{model0}

\subsection{\label{sec:couple}Coupling}
Now, we consider the coupling between a SNP and a MNP. As mentioned above, the process is divided into absorption and emission. Here, we define $x_1(t)$ and $x_2(t)$ as the displacements of oscillators of SNP and MNP, respectively.

Firstly, the absorption process is described as:
\begin{subequations}
\begin{align}
\ddot{x}_1+2 \beta_{a1} \dot{x}_1+\omega_{a1}^2 x_1-\gamma_2 \dot{x}_2-g_2^2 x_2=C_1 \mathrm{exp}(-\mathrm{i} \omega_{ex} t), \label{eq:couplea1} \\
\ddot{x}_2+2 \beta_{a2} \dot{x}_2+\omega_{a2}^2 x_2-\gamma_1 \dot{x}_1-g_1^2 x_1=C_2 \mathrm{exp}(-\mathrm{i} \omega_{ex} t). \label{eq:couplea2}
\end{align}
\label{eq:couplea}
\end{subequations}
Here, generally $C_1=C_2=C_0$, $\beta_{a1}$ and $\beta_{a2}$ are the damping coefficients of the SNP and the MNP in absorption process, respectively, $\omega_{a1}$ and $\omega_{a2}$ are the absorption resonance frequencies of the SNP and the MNP before coupling, respectively, $\gamma_j$ and $g_j$ ($j=1,2$) are the coupling coefficients, which are evaluated by:
\begin{equation}
\begin{aligned}
\gamma_j=\frac{N_j e^2}{2 \pi \varepsilon_0 m_e r^2 c},&\
 g_j^2=\frac{N_j e^2}{2 \pi \varepsilon_0 m_e r^3},
\\ j&=1,2,
\end{aligned}
\label{eq:g}
\end{equation}
where $N_1$ (or $N_s$) and $N_2$ (or $N_m$) are the effective electron numbers of the SNP and the MNP, respectively, and $r$ is the distance between the centers of the SNP and the MNP. The solutions of Eq. (\ref{eq:couplea}) are:
\begin{equation}
\begin{aligned}
x_j(t)&=A_j \mathrm{exp}(-\mathrm{i} \omega_{ex} t),
\\ A_1&=\frac{B_2+F_2}{B_1 B_2-F_1 F_2}C_0, \  A_2=\frac{B_1+F_1}{B_1 B_2-F_1 F_2}C_0,
\\ B_j&=-\omega_{ex}^2-2\mathrm{i}\beta_{aj}\omega_{ex}+\omega_{aj}^2,
\ F_j=g_j^2-\mathrm{i}\gamma_j \omega_{ex},
\\ j&=1,2,
\end{aligned}
\label{eq:sa}
\end{equation}

Secondly, the emission process is described as:
\begin{subequations}
\begin{align}
\ddot{x}_1+2 \beta_{e1} \dot{x}_1+\omega_{e1}^2 x_1-\gamma_2 \dot{x}_2-g_2^2 x_2=0, \label{eq:couplee1} \\
\ddot{x}_2+2 \beta_{e2} \dot{x}_2+\omega_{e2}^2 x_2-\gamma_1 \dot{x}_1-g_1^2 x_1=0, \label{eq:couplee2}
\end{align}
\label{eq:couplee}
\end{subequations}
where the initial conditions are: $x_j(0)=A_j$ and $\dot{x}_j=0$ ($j=1,2$). Here, $\beta_{e1}$ and $\beta_{e2}$ are the damping coefficients of the SNP and the MNP in emission process, respectively, and $\omega_{e1}$ and $\omega_{e2}$ are the emission resonance frequencies of the SNP and the MNP before coupling, respectively.
The solutions of Eq. (\ref{eq:couplee}) are similar to the ones in Ref. \mycite{couple0}, but with different initial conditions. We assume $x_1(t)=S\mathrm{exp}(\Omega t)$ and $x_2(t)=M\mathrm{exp}(\Omega t)$, and substitute them into Eq. (\ref{eq:couplee}) to obtain $\Omega$. Although $\Omega$ has analytic solutions, the expressions are too complex to be written here. Hence, we can rewrite the solutions of $\Omega$ in this form:
\begin{equation}
\begin{aligned}
\Omega_1^{\pm}=-\beta_1\pm\mathrm{i}\omega_1,
\\ \Omega_2^{\pm}=-\beta_2\pm\mathrm{i}\omega_2,
\end{aligned}
\label{eq:Omega}
\end{equation}
Thereby, combining with the initial conditions, the solutions of Eq. (\ref{eq:couplee}) can be written as:
\begin{equation}
\begin{aligned}
x_1(t)&=S_1\mathrm{exp}(\Omega_1^{-} t)+S_2\mathrm{exp}(\Omega_2^{-} t)
\\&+S_3\mathrm{exp}(\Omega_1^{+} t)+S_4\mathrm{exp}(\Omega_2^{+} t)
\\ &\cong S_1\mathrm{exp}(\Omega_1 t)+S_2\mathrm{exp}(\Omega_2 t),
\\ x_2(t)&=M_1\mathrm{exp}(\Omega_1^{-} t)+M_2\mathrm{exp}(\Omega_2^{-} t)
\\ &+M_3\mathrm{exp}(\Omega_1^{+} t)+M_4\mathrm{exp}(\Omega_2^{+} t)
\\ &\cong M_1\mathrm{exp}(\Omega_1 t)+M_2\mathrm{exp}(\Omega_2 t),
\end{aligned}
\label{eq:se}
\end{equation}
For the same reason as Eq. (\ref{eq:farw}), we omit the solutions marked with ``$+$'', i.e., $\Omega_1^{-}$ and $\Omega_2^{-}$ are retained. To make it simple, we define $\Omega_1=\Omega_1^{-}$ and $\Omega_2=\Omega_2^{-}$.
Therefore, the total emission far field in time domain can be written as:
\begin{equation}
\begin{aligned}
E(t)=&E_1(t)+E_2(t)
\\ \cong &\Omega_1^2 (K_s S_1+K_m M_1) \mathrm{exp}(\Omega_1 t)
\\ +&\Omega_2^2 (K_s S_2+K_m M_2) \mathrm{exp}(\Omega_2 t),
\\ := &A'_1 \mathrm{exp}(\Omega_1 t)+A'_2 \mathrm{exp}(\Omega_2 t),
\end{aligned}
\label{eq:emissiont}
\end{equation}
where $K_s=\frac{N_s e}{4 \pi \varepsilon_0 c^2 d }\propto N_s$ and $K_m=\frac{N_m e}{4 \pi \varepsilon_0 c^2 d }\propto N_m$.
The total emission intensity in frequency domain can be evaluated by:
\begin{equation}
I_{tot}(\omega)=\sum_{j=1}^2 |A'_j|^2 \frac{1-\mathrm{exp}(-2 \beta_j t_0)}{2 \beta_j t_0} \frac{\beta_j}{(\omega-\omega_j)^2+\beta_j^2}.
\label{eq:emissionw}
\end{equation}
Considering Fermi--Dirac distributions, the emission spectrum is tuned due to the electron temperature $T$. Therefore, the tuned emission intensity, i.e., the PL spectrum, can be evaluated by
\begin{equation}
I_{PL}(\omega)=I_{tot}(\omega) \frac{1}{1+\mathrm{exp}[\hbar (\omega-\omega_{ex})/(k_B T)]},
\label{eq:PL}
\end{equation}
where $\hbar$ is the reduced Planck constant and $k_B$ is the Boltzmann constant.

According to Eq. (\ref{eq:Omega}), there are two new modes, resulting in two peaks in the spectrum. One peak is related to the coupled SNP, the other peak is related to the coupled MNP, thereby, the PL spectrum can be divided into $I_{SNP}(\omega)$ and $I_{MNP}(\omega)$, where $I_{PL}=I_{SNP}+I_{MNP}$. Define the enhancement factor ($EF$) of SNP as:
\begin{equation}
EF_{SNP}=\frac{I_{SNP}^{(\mathrm{cp})}}{I_{SNP}^{(\mathrm{ind})}},
\label{eq:EF}
\end{equation}
where $I_{SNP}^{(\mathrm{cp})}$ and $I_{SNP}^{(\mathrm{ind})}$ stand for the peak intensities of the SNP mode of the coupled system and the individual SNP, respectively. We emphasize that for the coupled system, the SNP mode $I_{SNP}$ is emitted not only by the SNP but also by the MNP, because both the SNP and the MNP emit the two modes with corresponding amplitudes, and the total electric field is the coherent superposition of the two emissions. Therefore, the intensity of the SNP mode can be highly enhanced due to the fact that $N_2 \gg N_1$, i.e., the MNP help the SNP emit photons through the channel of the MNP with stronger intensity.

Notice that Eq. (\ref{eq:emissionw}) is similar to Eq. (19a) of Ref. \mycite{couple0}. However, the latter was employed to calculate the PL spectra from two coupled MNPs with the same $N$, thus same $g$ and $\gamma$; the former will be employed to calculate the PL spectra from coupled SNP and MNP which have different parameters, e.g., $N$ ($N_s$ and $N_m$) etc. In this work, the symmetry is broken, and novelty phenomena may be obtained.

\section{\label{sec:Results}Results and Discussions}
According to the model, the optical properties of the individual mode (SNP) and the coupling modes (SNP--MNP) are presented as following.

Firstly, the properties of the individual SNP is obtained.

\begin{figure}[tb]
\includegraphics[width=0.48\textwidth]{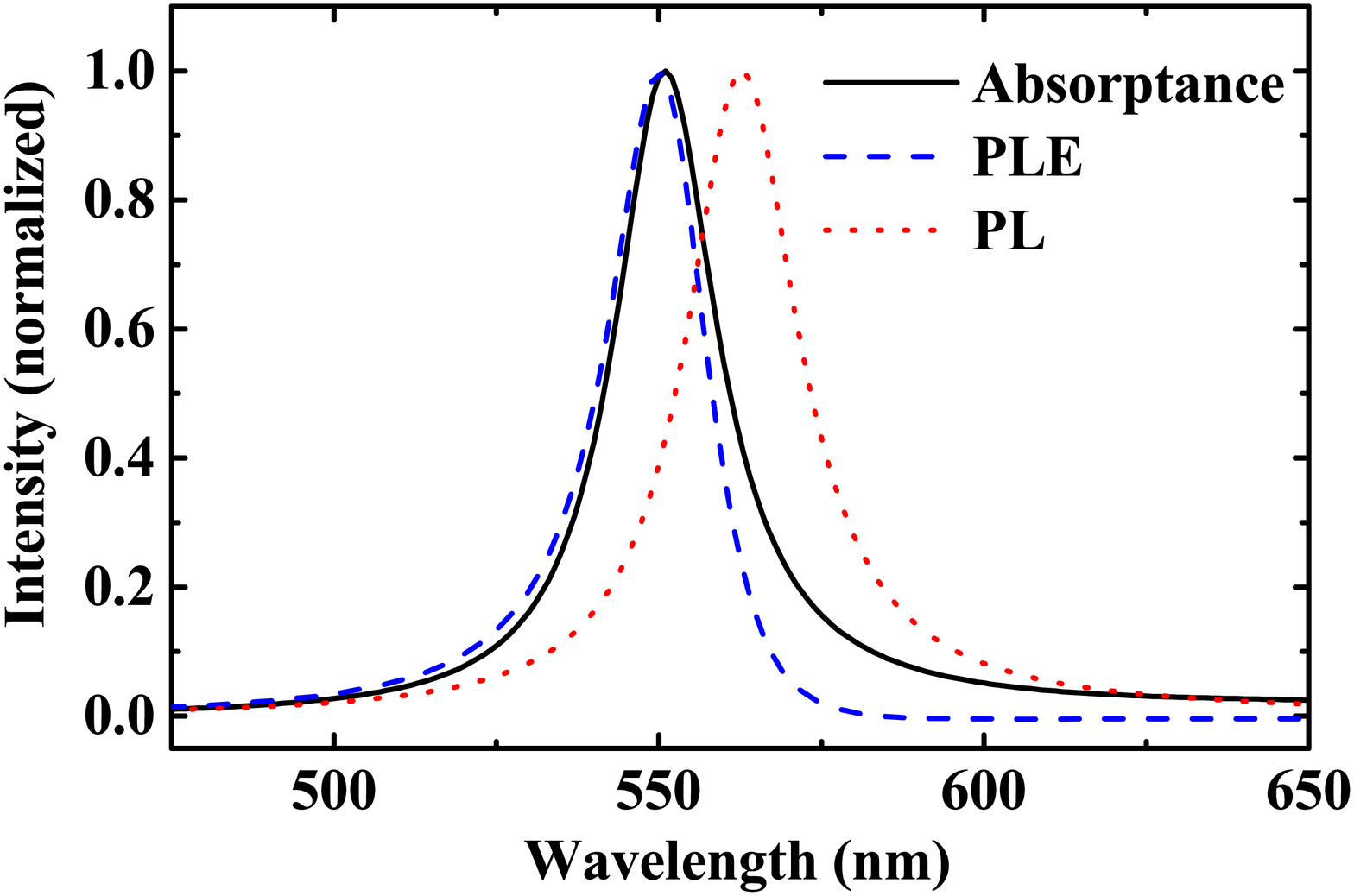}
\caption{\label{fig:PLE} Absorptance (solid black line), PLE (dashed blue line), and PL (dot red line), calculated from Eq. (\ref{eq:abs}), (\ref{eq:PLE}), and (\ref{eq:farw}), respectively. The parameters are: $\omega_a=2.251$ eV, $\beta_a=0.0396$ eV, $\omega_e=2.204$ eV, and $\beta_e=0.0408$ eV.
}
\end{figure}

Fig. \ref{fig:PLE} shows the calculated absorptance, PLE, and PL of an individual SNP with single absorption mode and single emission mode. The absorptance and PLE almost overlap for the larger frequencies, i.e., $\omega > \omega_a$, which is consistent with the experimental results of Ref. \mycite{PLE}. However, for smaller frequencies, i.e., $\omega < \omega_a$, the two curves do not overlap well. This is because when we use Eq. (\ref{eq:PLE}) to calculate PLE for $\omega < \omega_a$, the anti-Stokes emission of PL is not integrated, the lack of which indicates the difference between PLE and the absorptance. It is worth mentioning that a practical absorptance of a SNP has multiple modes covering a wide range of wavelengths from ultraviolet to visible range. Here, for simplicity, we just employ one mode for absorptance to demonstrate the optical properties.

Secondly, the coupled absorption process is investigated in detail according to Eq. (\ref{eq:sa}).

\begin{figure}[tb]
\includegraphics[width=0.48\textwidth]{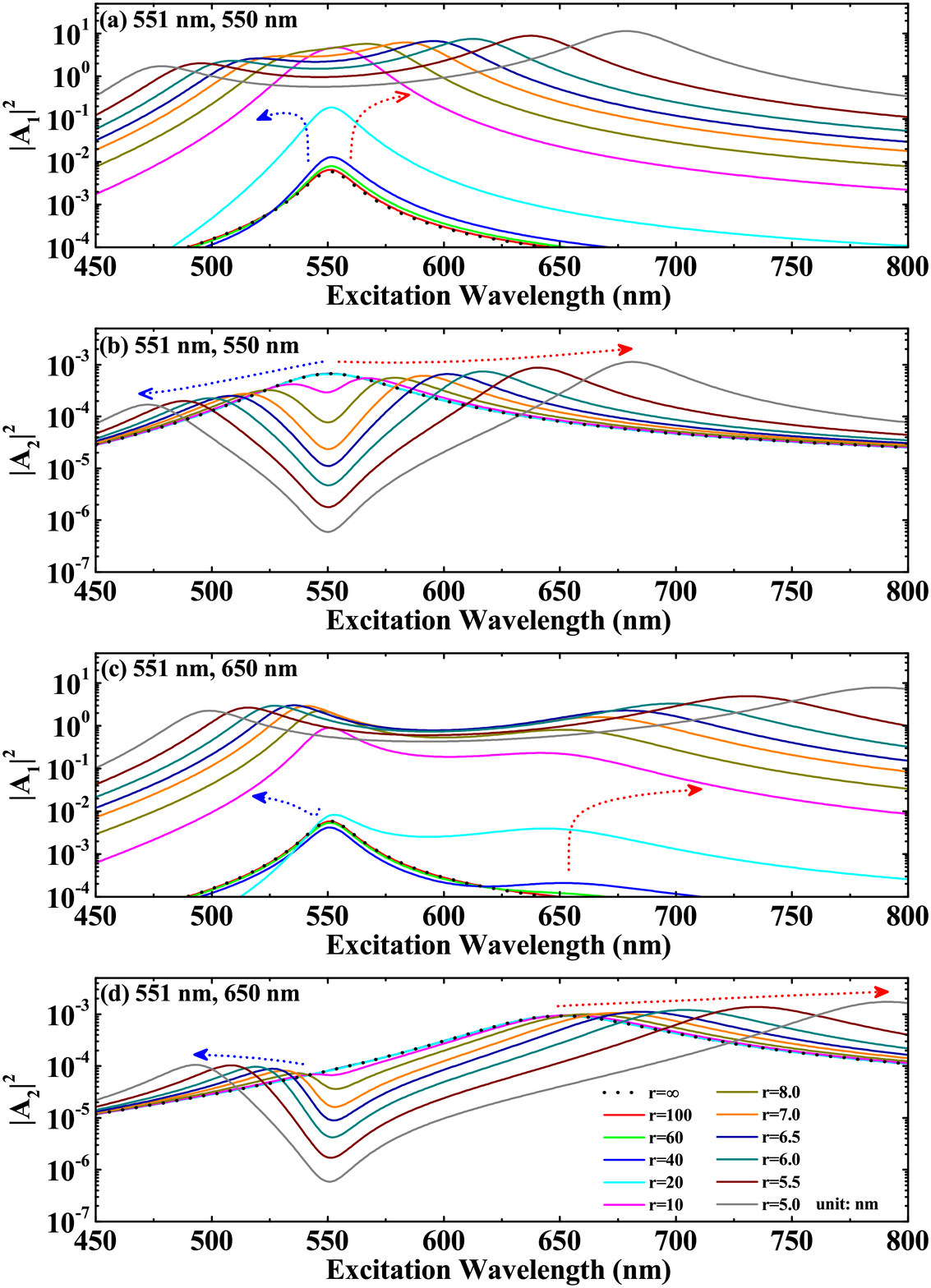}
\caption{\label{fig:Aex} The absorption amplitudes $|A_j|^2$ ($j=1,~2$) of coupled system with different distances $r$ as a function of the excitation wavelength, calculated form Eq. (\ref{eq:sa}). The dashed arrows stand for the trends of the blue and red shifts of the peaks as $r$ decreases. The legend is shown in (d) and is suitable for (a-d). The parameters are: (a-d) $N_1=10$, $N_2=10^5$; (a-d) $\omega_{a1}=2.251$ eV, $\beta_{a1}=0.0396$ eV (551 nm); (a,b) $\omega_{a2}=2.257$ eV, $\beta_{a2}=0.1175$ eV (550 nm); (c,d) $\omega_{a2}=1.911$ eV, $\beta_{a2}=0.1175$ eV (650 nm).
}
\end{figure}

Fig. \ref{fig:Aex} shows the absorption intensities ($|A_j|^2$) of the coupling system varying with excitation wavelength at different distances $r$.
We consider two situations, i.e., resonance coupling (Fig. \ref{fig:Aex}a-b) and non--resonance coupling (Fig. \ref{fig:Aex}c-d).
In resonance coupling, the original absorption peaks of the two NPs are close.
As the distance $r$ decreases, i.e., the coupling strength increases, at first ($r>10$ nm, weak coupling), the absorption amplitude of the SNP (Fig. \ref{fig:Aex}a, $|A_1|^2$) increases rapidly with almost no change in the line shape of the spectrum (one peak), while the one of the MNP (Fig. \ref{fig:Aex}b, $|A_2|^2$) remains almost unchanged in both intensity and line shape; and then ($r \le 10$ nm, strong coupling) both $|A_1|^2$ and $|A_2|^2$ appear two splitting peaks with the blue one's intensity decreasing and the red one's intensity increasing, and the splitting increases with the increase of the coupling strength.
In non--resonance coupling, the original absorption peaks of the two NPs are far apart.
As the distance $r$ decreases, at first ($r>10$ nm), the amplitude of the SNP (Fig. \ref{fig:Aex}c, $|A_1|^2$) decreases followed by the increase with the appearance of the MNP mode (around 650 nm) when $r<40$ nm, while the one of the MNP (Fig. \ref{fig:Aex}d, $|A_2|^2$) remains almost unchanged in both intensity and line shape; and then ($r \le 10$ nm) the two peaks of $|A_1|^2$ start to separate more with the blue one's intensity increasing followed by decreasing and the red one's intensity increasing, while the SNP mode starts to appear in $|A_2|^2$ with the two peaks separating more along with the increasing intensities.
The blue and red dashed arrows approximately represent the trends of the two modes with the increase of the coupling strength.

\begin{figure}[tb]
\includegraphics[width=0.48\textwidth]{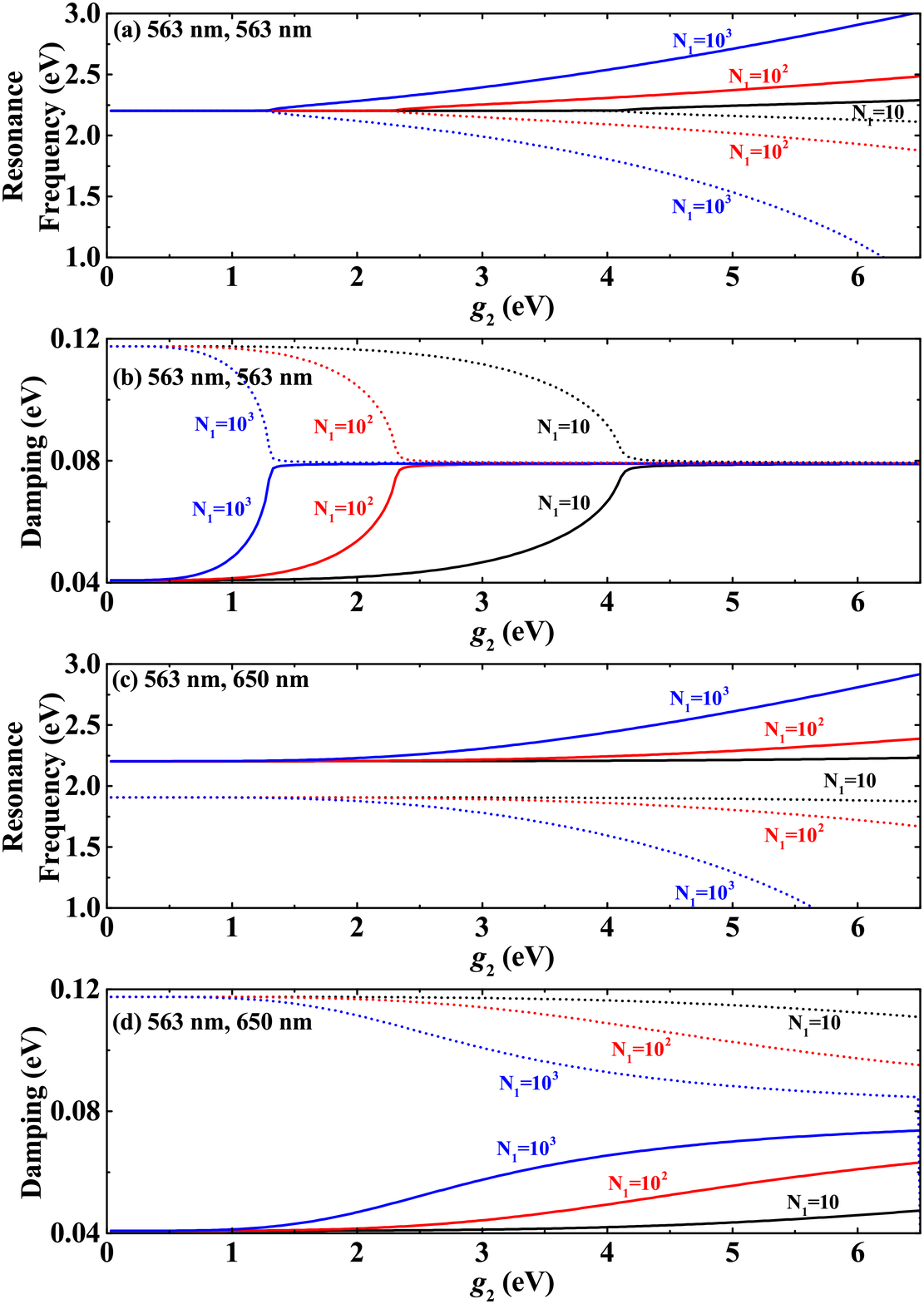}
\caption{\label{fig:modes} The resonance frequencies $\omega_1$ and $\omega_2$ (a,c) and damping coefficients $\beta_1$ and $\beta_2$ (b,d) of the coupled system as a function of $g_2$ with different $N_1$, calculated from Eq. (\ref{eq:Omega}). Black, red, and blue lines stand for $N_1=10$, $N_1=10^2$, and $N_1=10^3$, respectively. Solid and dashed lines with the same color stand for the two splitting modes. Here, (a-d) $N_2=10^5$, $\omega_{e1}=2.204$ eV, and $\beta_{e1}=0.0408 $ eV (563 nm); (a,b) $\omega_{e2}=2.257$ eV, and $\beta_{e2}=0.1175$ eV (563 nm); (c,d)
$\omega_{e2}=2.205$ eV, and $\beta_{e2}=0.1175$ eV (650 nm).
}
\end{figure}

When the coupling strength increases at weak coupling regime, why does the intensity of the SNP largely increase but the one of the MNP almost unchange? This is because the coupling strengths $g_1 \ll g_2$, i.e., the influence on the SNP from the MNP is much larger then the influence on the MNP from the SNP. When weakly coupled, the MNP is unaffected approximately, while the SNP is affected greatly. The reason why the splitting appears when strongly coupled will be discussed later in Fig. \ref{fig:modes}.

Notice that in Fig. \ref{fig:Aex}b and \ref{fig:Aex}d, the valley appears (at about 550 nm, corresponding to the resonance wavelength of the individual SNP) when the coupling strength is strong. Also notice that the absorption of the MNP is dominant compared with the one of the SNP due to the fact that $N_1 \ll N_2$. Therefore, the valley indicates that the absorption intensity of the system is extremely low when excited at the valley (550 nm). This is the phenomenon of electromagnetically induced transparency (EIT) introduced by classical mechanism. However, we should emphasize here that we only consider single mode for the absorption of the SNP, while the actual case is that the SNP has multi modes for the absorption, indicating the complicating coupling to achieve EIT.

Thirdly, the coupled emission process is investigated in detail.

Fig. \ref{fig:modes} shows the behaviors of the new generated modes of the coupled system for resonance coupling (Fig. \ref{fig:modes}a--b) and non--resonance coupling (Fig. \ref{fig:modes}c--d), respectively. The electron number of the MNP is kept as $N_2=10^5$. Hence, we use $g_2$ to represent one of the coupling strengths.
For a certain value of $N_1$, as $g_2$ increases, the coupled resonance frequencies ($\omega_1$, $\omega_2$) split along with the increasing of the splitting, as shown in Fig. \ref{fig:modes}a and \ref{fig:modes}c; while the coupled damping coefficients ($\beta_1$, $\beta_2$) approach and become the same at a large enough $g_2$, as shown in Fig \ref{fig:modes}b and \ref{fig:modes}d. That is, larger coupling strength ($g_1$ and $g_2$) results in larger splitting and smaller damping difference. On the other hand, for a certain value of $g_2$, as $N_1$ increases, the splitting of $\omega_1$ and $\omega_2$ increases; while the difference between $\beta_1$ and $\beta_2$ decreases and becomes zero at a large enough $g_2$. That is, larger $N_1$ results in larger coupling strength ($g_1$), thus larger splitting and smaller damping difference.

Furthermore, compared with the non--resonance coupling, the resonance coupling appears to be more splitting between the resonance frequencies and smaller differences between the damping coefficients in the same condition. For instance, at $g_2=2$ eV and $N_1=10^3$, the splitting of the resonance frequencies is about 163.2 and 55.42 meV for the resonance coupling and the non--resonance coupling, respectively; while the differences between the damping coefficients are about 0.5 and 64.7 meV for the resonance coupling and the non--resonance coupling, respectively. These phenomena indicate that the resonance coupling makes it easier to achieve stronger coupling.

\begin{figure}[tb]
\includegraphics[width=0.48\textwidth]{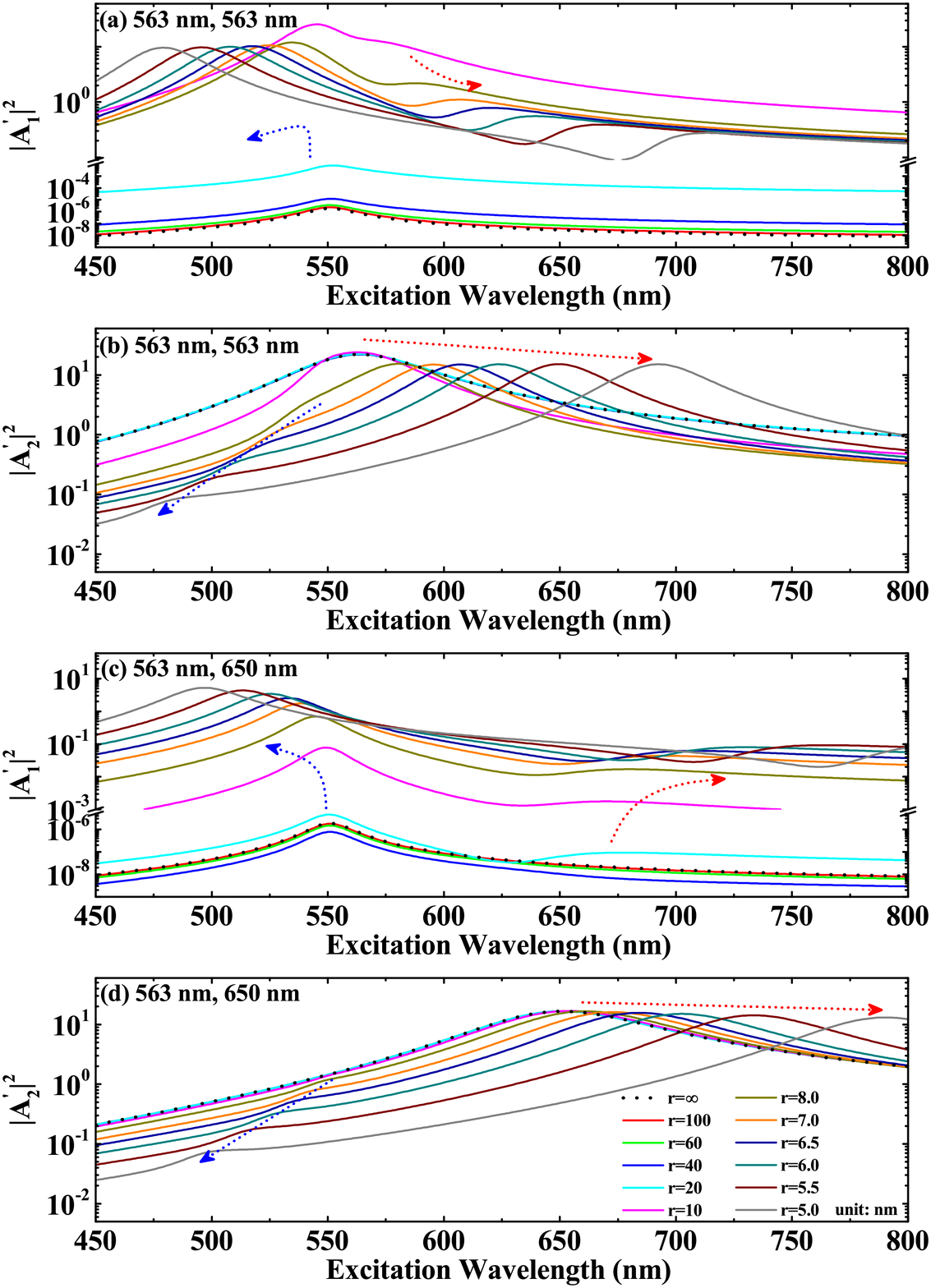}
\caption{\label{fig:Apex} The emission amplitudes $|A'|^2$ ($j=1,~2$) of the coupled system with different distances $r$ as a function of the excitation wavelength, calculated from Eq. (\ref{eq:emissiont}). The dashed arrows stand for the trends of the blue and red shifts of the peaks as $r$ decreases. The legend is shown in (d) and is suitable for (a-d). The parameters are: (a-d) $N_1=10$, $N_2=10^5$, $\omega_{a1}=$ eV, $\beta_{a1}=$ eV (551 nm), $\omega_{e1}=2.204$ eV, and $\beta_{e1}=0.0408$ eV (563 nm); (a,b) $\omega_{e2}=2.257$ eV, and $\beta_{e2}=0.1175$ eV (563 nm); (c,d)
$\omega_{e2}=2.205$ eV, and $\beta_{e2}=0.1175$ eV (650 nm).
}
\end{figure}

Fig. \ref{fig:Apex} shows the emission intensities ($|A'_j|^2$) of the coupling system varying with excitation wavelength at different distances $r$. Similarly, we consider the resonance coupling (Fig. \ref{fig:Apex}a--b) and non--resonance coupling (Fig. \ref{fig:Apex}c--d) situations.
In resonance coupling, as $r$ decreases, at first ($r>10$ nm), the emission amplitude of the SNP mode (Fig. \ref{fig:Apex}a, $|A'_1|^2$) increases rapidly with almost no change in the line shape of the spectrum, while the one of the MNP mode (Fig. \ref{fig:Apex}b, $|A'_2|^2$) remains almost unchanged in both intensity and line shape; and then ($r\le 10$ nm), both $|A'_1|^2$ and $|A'_2|^2$ appear two splitting peaks with the blue one's intensity decreasing and red one's intensity decreasing as well, and the splitting increases with the decrease of $r$.
In non--resonance coupling, as $r$ decreases, at first ($r>10$ nm), the amplitude of the SNP mode (Fig. \ref{fig:Apex}c, $|A'_1|^2$) decreases followed by the increase with the appearance of the MNP mode (around 650 nm), while the one of the MNP mode (Fig. \ref{fig:Apex}d, $|A'_2|^2$) remains almost unchanged in both intensity and line shape; and then ($r\le 10$ nm), the two peaks of $|A'_1|^2$ start to separate more with the blue one's intensity increasing and the red one's intensity increasing, while the SNP mode starts to appear in $|A'_2|^2$ with the two peaks separating more along with the blue one's intensity decreasing and the red one's intensity slightly decreasing.

Notice that these behaviors of the emission and absorption intensities as a function of $r$ and $\omega_{ex}$ are similar in general which has been shown in Fig. \ref{fig:Aex} and Fig. \ref{fig:Apex}. Therefore, the phenomena of the emission process can be explained in the same way as the absorption process, due to the fact that the two processes satisfy the similar equations, i.e., Eq. (\ref{eq:couplea}) and Eq. (\ref{eq:couplee}). The main differences are: (i) the trends of the red peaks of the splitting peaks are usually different; (ii) absorption process exists EIT, but emission  process does not. The differences are due to the different values of the quantities in Eq. (\ref{eq:couplea}) and Eq. (\ref{eq:couplee}), and the fact that the emission intensities $|A'_j|^2$ are affected by the absorption intensities $|A_j|^2$, which indicates that the whole process is complicated.

\begin{figure}[tb]
\includegraphics[width=0.48\textwidth]{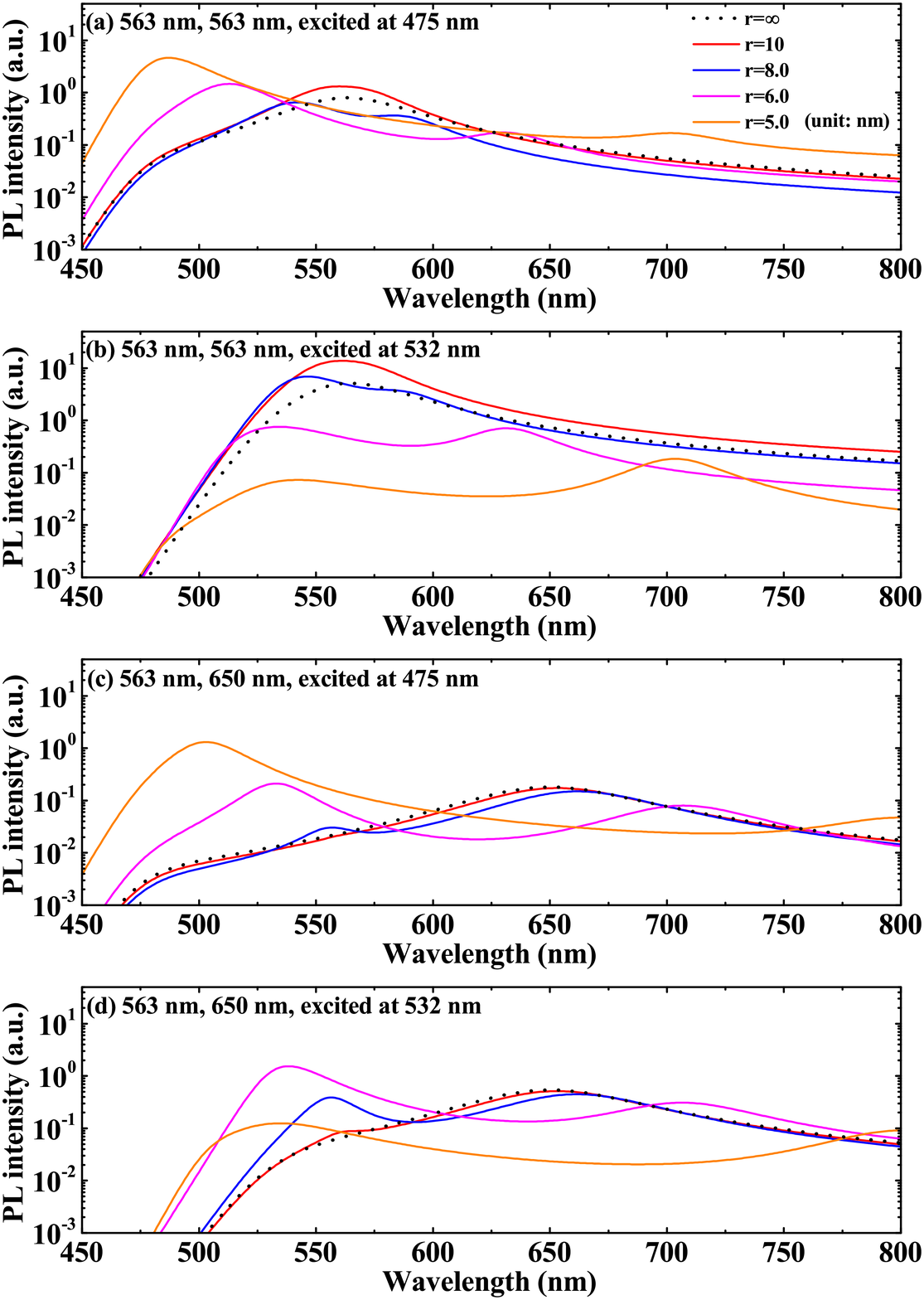}
\caption{\label{fig:IPL} The PL spectra of the coupled system with different $r$ and excited at 475 nm (a,c) and 532 nm (b,d), calculated from Eq. (\ref{eq:PL}). Here, T=500 K, $N_1=10$, $N_2=10^5$, and $\omega_{e1}$, $\beta_{e1}$, $\omega_{e2}$, and $\beta_{e2}$ are the same as Fig. \ref{fig:Apex}. The legend is shown in (a) and is suitable for (a-d).
}
\end{figure}

Fig. \ref{fig:IPL} shows the PL spectra of the coupling system with different $r$ excited at different wavelengths, also considering the resonance coupling (Fig. \ref{fig:IPL}a--b) and non--resonance coupling (Fig. \ref{fig:IPL}c--d).
In resonance coupling excited at 475 nm, with the decrease of $r$, the PL intensity firstly ($r\ge 10$ nm) increases with no evident splitting, and then ($r=8.0$ nm) decreases with splitting, followed by ($r=6.0,~5.0$ nm) the increase of the blue peak due to the fact that the blue peak is closer to the excited wavelength with smaller $r$ which is corresponding to resonance excitation. In resonance coupling excited at 532 nm, with the decrease of $r$, the PL intensity firstly ($r\ge 10$ nm) increases with no evident splitting, and then ($r<10$ nm) decreases with increasing splitting. Although the blue peak is close to 532 nm, the anti--Stokes emission of the PL spectra is restrained by the Fermi-Dirac distribution, resulting in the decrease of the intensity of the blue peak.
In non--resonance coupling excited at 475 nm, with the decrease of $r$, the PL intensity firstly ($r\ge 10$ nm) remains almost unchanged, and then ($r<10$ nm) increases in the blue peak with the increasing splitting due to the same reason as resonance coupling excited at 475 nm. In non--resonance coupling excited at 532 nm, with the decrease of $r$, the PL intensity firstly ($r\ge 10$ nm) remains almost unchanged, and then ($r=8.0,~6.0$ nm) increases in the blue peak with the increasing splitting, followed by ($r=5.0$ nm) the decrease of the blue peak due to the same reason as resonance coupling excited at 532 nm.

In general, the coupled PL spectra are affected by the excitation wavelength because one of the new generated modes will be close to the excitation wavelength with the increase of the coupling strength, indicating the resonance excitation and resulting in the enhancement of the spectra, as well as the weaken of the spectra caused by Fermi-Dirac distribution.

\begin{figure}[tb]
\includegraphics[width=0.48\textwidth]{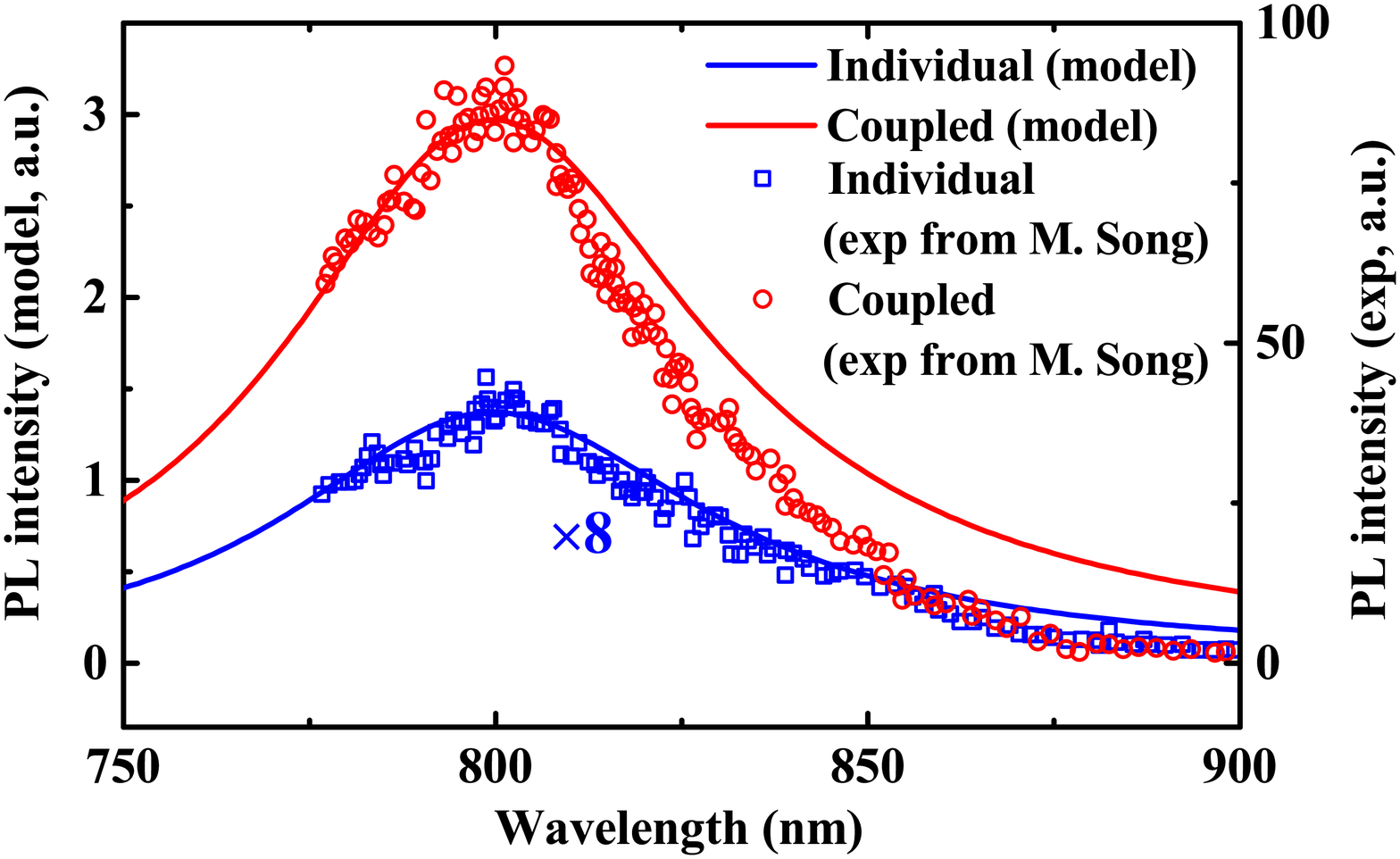}
\caption{\label{fig:IPLexp} The PL spectra of the individual SNP (blue) and the coupled SNP-MNP (red). Solid lines stand for the calculations from Eq. (\ref{eq:PL}), with the parameters: $r=18.9$ nm, T=500 K, $N_1=10$, $N_2=10^5$; $\omega_{a1}=1.60$ eV, $\beta_{a1}=0.06711$ eV (775 nm); $\omega_{e1}=1.5505$ eV, $\beta_{e1}=0.06711$ eV (800 nm); $\omega_{e2}=\omega_{a2}=2.350$ eV, and $\beta_{e2}=\beta_{a2}=0.43481$ eV (528 nm). Opened dots stand for the experimental data copied from M. Song \cite{Exp1}.
}
\end{figure}

To verify our model, a comparison with the experiments is necessary. Fig. \ref{fig:IPLexp} shows the comparisons between the calculations of our model and the experimental data from M. Song \cite{Exp1}. The blue open squares show the experimental PL spectra of the individual SNP (CdSeTe/ZnS QD) with single resonance wavelength at about 800 nm, copied from their paper \cite{Exp1}; while the blue curve shows the corresponding PL spectra calculated from our model with proper parameters. The shapes of these two agree well with each other. The red open circles show the PL spectra of coupled system, i.e., the QD coupled to a single Au microplate with the distance between them of $18\pm 1.9$ nm; while the red curve shows the corresponding PL spectra calculated from our model with the distance $r=18.9$ nm. The two spectra agree well in both the enhancement and the shape. Furthermore, we notice that the mode of the coupled system is almost the same as the mode of the individual SNP, indicating that the coupling is in the weak coupling regime, because there is no evident splitting in the coupling spectra. Although there is no splitting in weak coupling, there is an enhancement in the intensity of the SNP. The enhancement originates from the assistance of the MNP that emits the SNP modes through the channel of the MNP, which has been explained with Eq. (\ref{eq:EF}).

\begin{figure}[tb]
\includegraphics[width=0.48\textwidth]{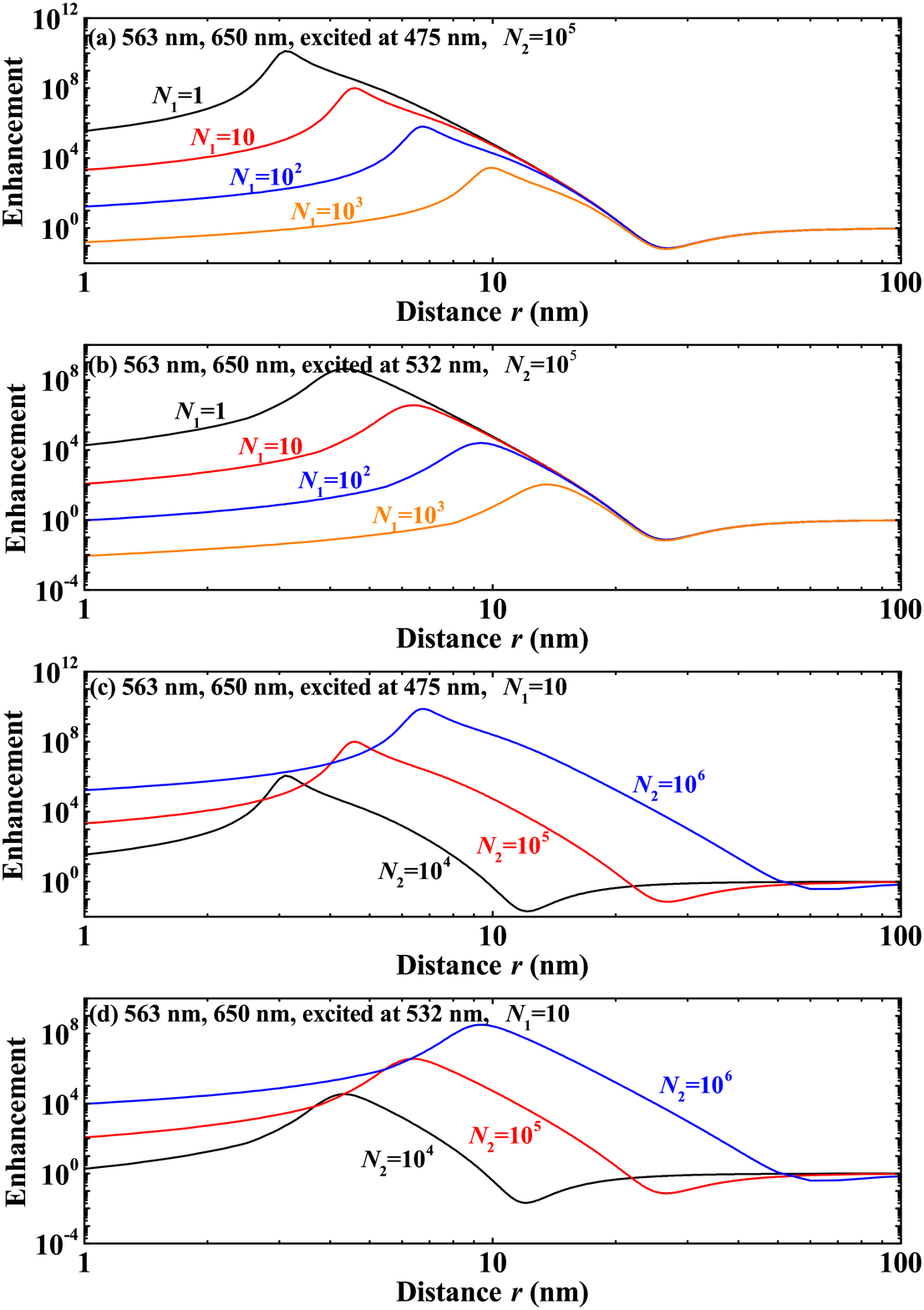}
\caption{\label{fig:EF} The $EF$ of the SNP as a function of $r$ with different $N_1$ (a,b) and $N_2$ (c,d), excited at 475 nm (a,c) and 532 nm (b,d), calculated from Eq. (\ref{eq:EF}). Here, T=500 K;  $\omega_{a1}=$ eV, $\beta_{a1}=$ eV (551 nm), $\omega_{e1}=2.204$ eV, and $\beta_{e1}=0.0408$ eV (563 nm); $\omega_{e2}=2.205$ eV, and $\beta_{e2}=0.1175$ eV (650 nm).
}
\end{figure}

Fig. \ref{fig:EF} shows the $EF$ of the SNP as a function of $r$ with different $N_1$ and $N_2$, and excited at different wavelengths, in the case of non--resonance coupling. In surface enhanced emission, non--resonance coupling is a more general case; moreover, resonance coupling have non--negligible background signals that are most from the MNP, which might submerge the weak signals from the SNP and hinders the detection. Therefore, here we only consider the general case, i.e., non--resonance coupling.
For all the cases in Fig. \ref{fig:EF}, as $r$ decreases, the $EF$ firstly decreases a little, and then increases rapidly to its maximum, followed by its decrease.

In Fig. \ref{fig:EF}a--b, $N_2$ is unchanged, and as $N_1$ increases, the maximum of the $EF$ decreases from about $10^{10}$ ($N_1=1$) to $10^3$ ($N_1=10^3$) excited at 475 nm, and from about $10^{8}$ ($N_1=1$) to $10^2$ ($N_1=10^3$) excited at 532 nm. Because the emission intensity of the individual SNP is proportional to $N_1^2$, the enhanced intensity of the SNP mode is proportional to $N_1^2\times EF$, resulting in the fact that the absolute intensity of the SNP mode of the coupled system is not that dependent on $N_1$. Therefore, for a certain MNP with certain $N_2$, although the $EF$ can reach very high when $N_1$ is enough low, the actually detected maximal signal of the SNP remains a certain order of magnitudes. Moreover, the maximum related distance $r$ increases as $N_1$ increases, indicating easier maximal coupling with larger $N_1$.

In Fig. \ref{fig:EF}c--d, $N_1$ is unchanged, and as $N_2$ increases, the maximum of the $EF$ increases from about $10^{6}$ ($N_2=10^4$) to $10^{10}$ ($N_2=10^6$) excited at 475 nm, and from about $10^{4}$ ($N_2=10^4$) to $10^{8}$ ($N_2=10^6$) excited at 532 nm. It indicates that larger $EF$ can be achieved by using MNP with larger $N_2$. This is because in the coupling, the SNP mode is most emitted by the MNP, which is proportional to $N_2$. Moreover, the maximum related distance $r$ increases as $N_2$ increases, indicating easier maximal coupling with larger $N_2$.

We notice that in Fig. \ref{fig:EF} the $EF$ excited at 475 nm is generally larger than the $EF$ excited at 532 nm, especially for the maximum; and there is a minimum for the $EF$ before it reaches the maximum as $r$ decreases. The corresponding case can be found in Fig. \ref{fig:Apex}c. When $r$ decreases, especially in the strong coupling regime, the blue--shift of the SNP mode slows down the increase of the intensity excited at 532 nm but speed up the increase of the intensity excited at 475 nm. On the other hand, the individual intensity (black dot line in Fig. \ref{fig:Apex}c) excited at 532 nm is larger than the one excited at 475 nm. The above two reasons make the results, i.e., the $EF$ excited at 475 nm is larger than the one excited at 532 nm. Furthermore, in Fig. \ref{fig:Apex}c, as $r$ decreases ($r>40$ nm), the intensity decreases to a minimum, corresponding to the minimum of the $EF$ in Fig. \ref{fig:EF}.

\section*{\label{sec:Conclusion}Conclusions}
In conclusion, we develop a classic model to reveal the optical properties of the individual SNP and as well as the plexciton, i.e., the coupling between a SNP and a MNP. Good agreements between our model and the published experiments verify the validity of our model. The model divides the whole process into absorption process and emission process, both of which are analyzed and investigated carefully. In the coupled system, the absorption properties reveal the splitting and the EIT in the spectra; the emission properties reveal the splitting in the spectra and the enhancement of the SNP. Also, the PL spectra are illustrated, and compared with the individual one, the spectral shapes are changed, i.e., modes split; and the intensities increase or decrease depending on the coupling strength. Moreover, the $EF$ is analyzed in detail, and varies with $N_1$, $N_2$, $\omega_{ex}$, and $r$. The maximum of the $EF$ can reach to the order of $10^{10}$ for a general case.
This work would be helpful to understanding the optical properties of plexciton, and the model is useful for related applications involving strongly coupled system of nanophotonics, such as strong coupling between the QDs and the MNPs.

%~\\ \indent% add an empty line
\section*{\label{sec:Acknowldegment}Acknowledgment}
This work was supported by the Fundamental Research Funds for the Central Universities (Grant No. FRF-TP-20-075A1).

\section*{Disclosures}
The authors declare no conflicts of interest.

\section*{Data availability}
The data that support the findings of this study are available from the corresponding author upon reasonable request.

\section*{\label{sec:Ref}References}
%\bibliography{viscosity_Cheng}
%\balance
\bibliography{PE_Cheng}

\end{document}